\documentclass[12pt]{article}
\usepackage{amsfonts}

\newfont{\twelvemsb}{msbm10 scaled\magstep1}
\newfont{\eightmsb}{msbm8}
\newfam\msbfam
\textfont\msbfam=\twelvemsb \scriptfont\msbfam=\eightmsb
\catcode`\@=11
\def\Bbb{\ifmmode\let\next\Bbb@\else
\def\next{\errmessage{Use \string\Bbb\space only in math mode}}\fi\next}
\def\Bbb@#1{{\fam\msbfam{{#1}}}}

\begin{document}
\sloppy
\renewcommand{\thefootnote}{\fnsymbol{footnote}}
\newpage
\setcounter{page}{1} \vspace{0.7cm}
\begin{flushright}
January 2006\\
LAPTH - 1139/06
\end{flushright}
\vspace*{1cm}
\begin{center}
{\bf  The elliptic scattering theory of the 1/2-XYZ and higher order Deformed Virasoro Algebras}\\
\vspace{1.8cm} {\large Davide Fioravanti$^a$ and \ Marco Rossi$^b$
\footnote{E-mail: Davide.Fioravanti@bo.infn.it, rossi@lapp.in2p3.fr}}\\
\vspace{.5cm} $^a${\em Department of Mathematics, University of York (UK),\\
Department of Physics and Mathematics, University of Trento (Italy),\\
INFN and Department of Physics, University of Bologna (Italy)} \footnote{The last one from December 19-th 2005 onward.}\\
\vspace{.3cm} $^b${\em Lapth \footnote{UMR 5108 du CNRS,
associ\'ee \`a
    l'Universit\'e de Savoie.}, 9 Chemin de Bellevue, BP 110, F-74941 Annecy-le-Vieux Cedex, France} \\
\end{center}
\renewcommand{\thefootnote}{\arabic{footnote}}
\setcounter{footnote}{0}
\begin{abstract}
{\noindent Bound} state excitations of the spin 1/2-XYZ model are
considered inside the Bethe Ansatz framework by exploiting the
equivalent Non-Linear Integral Equations. Of course, these bound
states go to the sine-Gordon breathers in the suitable limit and
therefore the scattering factors between them are explicitly
computed by inspecting the corresponding Non-Linear Integral
Equations. As a consequence, abstracting from the physical model
the Zamolodchikov-Faddeev algebra of two $n$-th elliptic breathers
defines a tower of $n$-order Deformed Virasoro Algebras,
reproducing the $n=1$ case the usual well-known algebra of
Shiraishi-Kubo-Awata-Odake \cite{SKAO}.
\end{abstract}

\vspace{1cm} {\centerline {\sl To our friend Daniel, ad memoriam}}

\vspace{2cm} {\noindent {\it Keywords}}: Deformed Virasoro
Algebra, Integrability; Non-Linear Integral Equation; S-matrix;
1/2-XYZ spin chain.

\newpage

\section{Prelude}
At this stage, integrable one-dimensional quantum spin chains are
a widely studied subject in theoretical physics. In fact, they are
interesting models in themselves as integrable (qualitative)
description of interactions between microscopic magnets in solid
state physics phenomena and, besides, they may be thought of as
(lattice) regularisations of two dimensional quantum field
theories (cf. for instance \cite{LUT}, \cite{DDV}): in this
respect, they opened the road of the application of techniques,
like the Algebraic Bethe Ansatz, which would be otherwise
considerably difficult if applied directly to the infinite degrees
of freedom case. Finally, recent and surprising findings (see
\cite{MZ} and the subsequent development) showed that hamiltonians
of some integrable spin chains coincide with the mixing matrices
of gauge invariant operators of ${\cal N}=4$ super Yang-Mills
theories, opening a new way to test the AdS/CFT correspondence.

Given this growing interest in the subject, we started by studying
the most general one dimensional spin 1/2 chain, the so-called
1/2-XYZ, which describes spin 1/2 degrees of freedom on a lattice
with anisotropic nearest-neighbour interaction. In the infinite
volume limit and with finite lattice spacing (thermodynamic
limit), a suitable scaling of the elliptic nome allowed the
authors of \cite{LUT} to obtain an integrable quantum field theory
in the two dimensional plane, the sine-Gordon model. Therefore, it
is reasonable to think of the 1/2-XYZ chain (in the thermodynamic
limit) as an elliptic deformation of the sine-Gordon field theory.
Moreover, in a recent paper \cite{FRXYZ} we proved that the same
conclusion holds if we also allow the lattice spacing to go to
zero in such a way to keep finite the lattice length on a
circumference (cylinder geometry). In fact, starting from the
Algebraic Bethe Ansatz equations we wrote down the equivalent
Non-Linear Integral Equations (NLIEs) describing the
anti-ferromagnetic ground state, its first excitations, -- i.e.
spin waves propagating along the chain --, and bound states
thereof, which turn out to be labelled by a positive integer $n$
\cite{FRXYZ}. These excitations become in the sine-Gordon limit
the solitons/antisolitons and breathers respectively. Therefore,
we may think of them as elliptic deformed counterparts of
solitons/antisolitons and breathers, and definitely borrow these
names from the field theory parlance. By inspecting the NLIE in
the thermodynamic limit, we started to compute scattering factors
between excitations \cite{FRXYZ}. Our aim was to reveal, within a
clear and unique framework, the elliptic scattering factors which
had been already proposed \cite{ZAM,LUK,MP}, without however
transparent connection to physical models. In this respect, the
scattering matrix between excitations in the repulsive regime --
elliptic solitons and antisolitons -- was computed and showed to
coincide with Zamolodchikov's S-matrix \cite{ZAM}. The elliptic
nome of such a matrix was found shifted with respect to the one
appearing in the parameters defining the XYZ chain, thereby
proving the so-called Smirnov's conjecture. On the other hand, in
the attractive regime the scattering factor between two first
($n=1$) elliptic breathers was also computed \cite{FRXYZ} and
showed to coincide on one side with the Lukyanov-Mussardo-Penati
S-factor \cite{LUK,MP}, on the other side with the elliptic
structure function of the Deformed Virasoro Algebra \cite{SKAO}.

In this paper we will continue the calculation of S-matrices by
computing all the possible scattering factors between two $n$ and
$m$ elliptic breathers, with $n,m \geq 1$. As expected, these
factors are elliptic deformations of the corresponding ones
between sine-Gordon breathers \cite{ZZ}. From the mathematical
point of view they may be used, when $n=m$, to define an order $n$
Deformed Virasoro Algebra (DVA). For the well-known DVA (order
$n=1$) \cite{SKAO} is obtained by quantizing the
Frenkel-Reshetikhin Poisson bracket \cite{FrRe} and all the other
higher order ones ($n>1$) may be thought of in the same manner.

\section{An outlook on the theoretical background}
\setcounter{equation}{0} The spin 1/2-XYZ model with periodic
boundary conditions is a spin chain with hamiltonian, written in
terms of Pauli matrices $\sigma^{x,y,z}$,
\begin{equation}
{\cal H}=\frac {1}{2}\sum _{n=1}^N(J_x \sigma _n^x \sigma
_{n+1}^x+J_y\sigma _n^y \sigma _{n+1}^y+J_z \sigma _n^z \sigma
_{n+1}^z) \, . \label {Hamilt}
\end{equation}
Here $N$ is the number of lattice sites and because of the
periodicity the site $N+1$ is identified with the site $1$. The
three (real) coupling constants $J_x$, $J_y$ and $J_z$ may be
reparametrised (up to an overall multiplicative constant) in terms
of elliptic functions with nome ${\mbox {exp}}\left (-\pi \frac
{{\bf K}'}{{\bf K}}\right)$:
\begin{equation}
J_x=1+{\bf k}\, {\mbox {sn}}^2 2\eta \, , \quad J_y=1-{\bf k}\,
{\mbox {sn}}^2 2\eta \, , \quad J_z={\mbox {cn}}2\eta \, {\mbox
{dn}} 2\eta \, , \label {Jpar}
\end{equation}
where $\eta$ is real.
\medskip
As for the notations used in this paper we address the reader, for
instance, to the Section 2 of \cite{FRXYZ}. As in \cite {FRXYZ},
we constrain our analysis to the disordered regime, i.e.
\begin{eqnarray}
&& 0<{\mbox {exp}}\left (-\pi \frac
{{\bf K}'}{{\bf K}}\right)<1  \, , \nonumber \\
&& 0<\eta < {\bf K} \, , \label {regime}
\end{eqnarray}
which are equivalent to ${\bf k}>0$, ${\bf K}>0$, ${\bf K}'>0$ or
$|J_z|<J_y<J_x$. We also distinguish between the repulsive regime
$0<\eta < \frac {\bf K}{2}$, and the attractive regime $\frac {\bf
K}{2} <\eta < {\bf K}$.

\medskip

The spin 1/2-XYZ chain enjoys many interesting limit cases. If
${\bf K}' \rightarrow \infty$, which implies ${\bf K}\rightarrow
\frac {\pi}{2}$, we have $J_x \rightarrow J_y \rightarrow 1$, $J_z
\rightarrow \cos 2 \eta$, so that we recover the XXZ chain in the
massless anti-ferromagnetic regime. If ${\bf K}\rightarrow \infty$
(${\bf K}'\rightarrow \frac {\pi}{2}$), we discover that
$J_y\rightarrow J_z\rightarrow \frac {1}{\cosh ^2 2\eta}$, $J_x
\rightarrow 1+\tanh ^2 2\eta$, which means that we obtain the XXZ
chain in the anti-ferromagnetic massive regime. As mentioned in
the prelude, particular attention should be payed to the double
scaling limit of site number and coupling constant
\begin{equation}
N\rightarrow \infty \, , \quad {\bf K}'= \frac {8 \eta}{\pi}\log N
+ {\mbox {finite terms}} \, ,
\end{equation}
on the cylinder of constant circumference $R$ and lattice spacing
$\Delta =\frac{R}{N}\rightarrow 0$. In this case we obtain the
sine-Gordon model on a cylinder of circumference $R$ with coupling
constant $\beta ^2=1-\frac {2\eta}{\pi}$, i.e. with Lagrangian
${\cal L}=\frac {1}{2}(\partial \phi )^2+\frac {m_0^2}{8\pi \beta
^2}(\cos \sqrt {8\pi}\beta \phi -1)$ \cite{FRXYZ}. Eventually, if
even $R$ goes to infinity, we gain the plane geometry, where the
famous results of \cite{LUT} apply.

\medskip

The spectrum of the spin 1/2-XYZ model was found in \cite{TF} by
means of the Algebraic Bethe Ansatz. In particular, the Bethe
equations read
\begin{equation}
\left [ \frac {H(i\alpha _j +\eta)\Theta (i\alpha _j +\eta)}
{H(i\alpha _j -\eta)\Theta (i\alpha _j -\eta)}\right]^N= -e^{-4\pi
i \nu \frac {\eta}{2{\bf K}}}\prod _{k=1}^M \frac{ H(i\alpha _j -
i\alpha _k +2\eta )\Theta (i\alpha _j - i\alpha _k +2\eta )} {
H(i\alpha _j - i\alpha _k -2\eta )\Theta (i\alpha _j - i\alpha _k
-2\eta )} \, , \label{bethe}
\end{equation}
where $j=1,...,M$ and $\nu$ is an integer. These equations
(\ref{bethe}) are valid when
\begin{equation}
m_1\eta = 2m_2 {\bf K}  \, ,  \label {relpar}
\end{equation}
where $m_1$ and $m_2$ are integers, and when $2M=N$, mod $m_1$.
The corresponding eigenvalues of the transfer matrix $\Lambda
_N(\alpha)$ may be written in terms of the Bethe roots as
\begin{eqnarray}
\frac {\Lambda_N(\alpha)}{\Theta (0)^N}&=&e^{2\pi i \nu \frac
{\eta}{2{\bf K}}}H(i\alpha +\eta)^N \Theta (i\alpha +\eta)^N\prod
_{j=1}^{M}\frac {H(i\alpha _j - i\alpha +2\eta)\Theta (i\alpha _j
-i\alpha
+2\eta)}{H(i\alpha _j-i\alpha)\Theta (i\alpha _j-i\alpha)}+\nonumber \\
\label {traeig0}  \\
&+&e^{-2\pi i \nu \frac {\eta}{2{\bf K}}}H(i\alpha -\eta)^N \Theta
(i\alpha -\eta)^N\prod _{j=1}^{M}\frac {H(i\alpha - i\alpha _j
+2\eta)\Theta (i\alpha -i\alpha _j+2\eta)}{H(i\alpha -i\alpha
_j)\Theta (i\alpha -i\alpha _j)} \, . \nonumber
\end{eqnarray}
While writing (\ref {bethe}, \ref {traeig0}) we borrowed the
notations from \cite{TF}: in particular, all the elliptic
functions have nome ${\mbox {exp}}\left (-\pi \frac {{\bf
K}'}{{\bf K}}\right)$.

\section{The Non-Linear Integral Equation}
\setcounter{equation}{0} In \cite {FRXYZ} we described the vacuum
and the first excitations of the spin 1/2-XYZ chain upon
transforming the Bethe equations into an equivalent NLIE, one for
each state. In Algebraic Bethe Ansatz language the vacuum is given
by $\nu =0$ and by filling the real interval $]-\frac {{\bf
K}'{\bf K}}{2\eta} \, , \frac {{\bf K}'{\bf K}}{2\eta}[$ with
Bethe roots. Excitations over the vacuum are expressed by $N_h$
(real) holes, $N_c$ close (complex) pairs, $N_w$ wide (complex)
pairs and $N_{sc}$ (complex) self-conjugated roots, the vacuum
obviously corresponding to $N_h=N_c=N_w=N_{sc}=0$ (cf.
\cite{FRXYZ} for details). In fact, any of these states may be
described in a compact way by means of the following NLIE for the
unknown function $\tilde Z_N$
\begin{eqnarray}
\tilde Z_N( \theta)&=&N \tilde F(\theta) +\sum _{k=1}^{N_h}\tilde
\chi (\theta -h_k) -\sum _{k=1}^{N_c}\tilde \chi ( \theta -c_k)
-\sum _{k=1}^{N_w}\tilde \chi _{II}( \theta -w_k) - \nonumber \\
&-& \sum _{k=1}^{N_{sc}}\tilde \chi _{II}(\theta -s_k)+2 \int
_{-\frac {{\bf K}{\bf K}'}{2\eta}}^{\frac {{\bf K}{\bf
K}'}{2\eta}}d \eta  \, \tilde G(\theta-\eta)\, {\mbox {Im}}\ln
\left [1+e^{i\tilde Z_N(\eta+i0)}\right] \, , \label{ddvex}
\end{eqnarray}
where we assume for granted from \cite{FRXYZ} the definitions of
the functions of the renormalised rapidity $\theta$, $\tilde
Z_N(\theta)=Z_N \left (\frac {\eta}{\bf K} \theta \right )$
(similarly for $F$ and $\chi$) and $\tilde G (\theta)=\frac
{\eta}{\bf K}G\left (\frac {\eta}{\bf K} \theta \right )$.
Nonetheless, we recall for the lazy reader the explicit formul\ae
\, for {\it the forcing term}, {\it the kernel} and {\it the
scattering phase}
\begin{equation}
\tilde F(\theta)=\sum _{\stackrel {n=-\infty}{n \not=
0}}^{+\infty} \frac {i}{n}\frac {\sinh \frac {2n({\bf
K}-\eta)\pi}{{\bf K}'} - \left (1-\frac {\eta}{\bf K}\right)\cos
n\pi \sinh \frac {2n{\bf K}\pi}{{\bf K}'}}{\sinh \frac {2n{\bf
K}\pi}{{\bf K}'}+\sinh \frac {2n({\bf K}-2\eta)\pi}{{\bf
K}'}}e^{-2i \frac {n \pi \eta \theta}{{\bf K}{\bf K}'}} \, ,\label
{force}
\end{equation}
\begin{equation}
\tilde G(\theta)=\frac {\eta}{{\bf K}{\bf K}'}\sum _{\stackrel
{n=-\infty}{n \not= 0}}^{+\infty} \frac {\sinh \frac {2n({\bf
K}-2\eta)\pi}{{\bf K}'}}{2\sinh \frac {2n({\bf K}-\eta)\pi}{{\bf
K}'} \cosh \frac {2n\eta \pi}{{\bf K}'}}  e^{-2i \frac {n \pi \eta
\theta}{{\bf K}{\bf K}'}} \, , \label {G}
\end{equation}
\begin{eqnarray}
&&\tilde \chi (\theta)=\int _0^{ \theta} dx
\, 2 \pi \, \tilde G(x)=\nonumber \\
&=&i\sum _{\stackrel {n=-\infty}{n \not= 0}}^{+\infty} \frac
{\sinh \frac {2n({\bf K}-2\eta)\pi}{{\bf K}'}}{2\sinh \frac
{2n({\bf K}-\eta)\pi}{{\bf K}'} \cosh \frac {2n\eta \pi}{{\bf
K}'}}  \frac {e^{-2i \frac {n \pi \eta \theta}{{\bf K}{\bf
K}'}}}{n} \, , \quad |{\mbox {Im}}\theta |<2\frac {{\bf
K}^2}{\eta}-2{\bf K} \, . \label{chi}
\end{eqnarray}
Besides, we required reference to the so-called second
determination of $\tilde \chi$, $\tilde \chi _{II}$, in that wide
roots lie, by definition, outside the strip $|{\mbox {Im}}\theta
|<{\mbox {min}} \{ 2 {\bf K} \, , \, 2{\bf K}^2/\eta -2{\bf K}\}$,
where formula (\ref {chi}) rigorously holds. Actually, here we
only need the second determination in the attractive regime $\frac
{\bf K}{2} <\eta < {\bf K}$
\begin{equation}
\tilde \chi _{II}(\theta)=\tilde \chi (\theta)-\tilde \chi (\theta
- 2i ({\bf K}^2/\eta - {\bf K}) \, {\mbox {sgn}}{\mbox
{Im}}\theta) \, , \label{chi2}
\end{equation}
or better, spitting the upper from the lower band, we may
equivalently introduce the two functions
\begin{equation}
\tilde \chi _{II}^{(\epsilon)}(\theta)=\tilde \chi _{II}(\theta)
\quad {\mbox {when}} \, \, \epsilon \, {\mbox {Im}}\, \theta
>2{\bf K}^2/\eta -2{\bf K} \, , \quad \epsilon =\pm 1 \, .
\label{chi2eps}
\end{equation}
It is convenient to re-express (\ref {chi2}) in terms of the
multiplicative independent variables
\begin{equation}
x={\mbox {exp}}\left (-\frac {2i\pi \eta \theta}{{\bf K}{\bf
K}'}\right ) \, , \quad q={\mbox {exp}}\left (-\frac {4\pi {\bf
K}}{{\bf K}'} \right ) \, , \quad p={\mbox {exp}}\left (-\frac
{4\pi \eta}{{\bf K}'} \right ) \, ,
\end{equation}
which naturally lead to the definitions ($|a|<1$)
\begin{equation}
(x;a)=\prod _{s=0}^{\infty} (1-xa^s) \, ,\quad  \Theta
_a(x)=(x;a)(ax^{-1};a)(a;a) \, . \label{thetafunc}
\end{equation}
The latter satisfies the multiplicative properties
\begin{equation}
\Theta _a(ax)=\Theta _a(x^{-1})=-x^{-1}\Theta _a(x) \, .
\label{thetaprop}
\end{equation}
Thanks to the definitions (\ref {thetafunc}) we may write
(\ref{chi2eps}) explicitly
\begin{equation}
\tilde \chi _{II}^{(\epsilon)}(\theta)=i \ln \left [ \frac {\Theta
_{p^2} (xq^{\frac {\epsilon+1}{2}}p^{-\frac
{\epsilon+1}{2}})\Theta _{p^2} (x^{-1}q^{\frac
{1-\epsilon}{2}}p^{-\frac {1-\epsilon}{2}})} {\Theta _{p^2}
(xq^{\frac {\epsilon+1}{2}}p^{-\frac {\epsilon-1}{2}})\Theta
_{p^2} (x^{-1}q^{\frac {1-\epsilon}{2}}p^{\frac {\epsilon+1}{2}})}
\right ] ^\epsilon \, . \label {chi22}
\end{equation}

\section{The bound states, i.e. the elliptic breathers}
\setcounter{equation}{0} In the attractive regime ${\bf K}/2 <\eta
<{\bf K}$ a soliton and an antisoliton may form a bound state,
which may be named -- in analogy with the sine-Gordon theory --
elliptic breather. It is represented as Bethe root configuration
by adding to the Fermi-Dirac sea of the vacuum a string of complex
(conjugated) roots \footnote{This string was the simplest one,
namely a pair, for the low-lying breather of \cite{FRXYZ}; besides
the soliton/antisoliton and breather excitation exist and are
described by the regular string values below only in the
thermodynamic limit $N\rightarrow\infty$.}. After remembering
\cite{LUT} that in the thermodynamic limit $N\rightarrow\infty$
the breather $B_n(\theta)$ with (real) rapidity $\theta$ can exist
only if the integer $n$ satisfies
\begin{equation}
\frac {n}{n+1} {\bf K}< \eta  \, , \label {cond}
\end{equation}
we must consider separately the two cases of $n$ even and odd: the
breather $B_{2k+2}(\theta)$ is described by the string ($\theta$
real)
\begin{equation}
\theta \pm i{\bf K}\pm i{\bf K} 2h \left (1-\frac {{\bf K}}{\eta}
\right) \, , \quad h=0,1,...,k \, ,
\end{equation}
while the breather $B_{2k+1}(\theta)$ by the string
\begin{equation}
\theta +i \frac {{\bf K}^2}{\eta} \, , \quad \theta \pm i{\bf
K}\pm i{\bf K}(2h-1) \left (1-\frac {{\bf K}}{\eta} \right) \, ,
\quad h=1,...,k \, .
\end{equation}
As a consequence of the constraint (\ref{cond}) the previous
strings contain wide roots only, i.e. they are entirely in the
second analyticity region of $\tilde Z_N (\theta)$. An explanation
of this fact will be given in a forthcoming paper \cite{35} and
will justify the constraint (\ref{cond}) itself.

We are now ready to study the collision two breathers of the XYZ
spectrum. Besides the clear interest as an example of
non-relativistic factorised (integrable) theory, this scattering
acquires a peculiar r\^ole in a mathematical perspective, since we
showed in a previous publication \cite{FRXYZ} that in the case of
two low-lying ($n=1$) breathers it furnishes, -- as
Zamolodchikov-Faddeev algebra --, the Deformed Virasoro Algebra
(DVA) of \cite{SKAO}. In order to obtain the asymptotic states, we
shall consider the thermodynamic ($N\rightarrow \infty$) limit of
the chain.

\section{Scattering factors between two elliptic breathers}
\setcounter{equation}{0}

When the number of lattice sites $N$ goes to infinity, the NLIE
undergoes a significant simplification in that the convolution
term can be neglected with respect to the other terms
\cite{DDV,FMQR,BOL}. Hence, when $N\rightarrow \infty$ the NLIE
(\ref{ddvex}), specified for the configuration containing an
elliptic breather $B_{2k'+2}(\theta _1)$ and another breather
$B_{2k+2}(\theta _2)$, takes on the simplified form
\begin{eqnarray}
\tilde Z_{N}(\theta)&=&N\tilde F( \theta)-\sum _{h'=0}^{k'} \sum
_{\epsilon '=\pm 1}\tilde \chi _{II}^{(\epsilon ')}\left (\theta
-\theta _1 + i \epsilon ' {\bf K} + i\epsilon ' {\bf K}2h' \left
(1-\frac {{\bf K}}{\eta}\right) \right ) - \nonumber \\
&-&\sum _{h=0}^k\sum _{\epsilon =\pm 1} \tilde \chi
_{II}^{(\epsilon)}\left (\theta -\theta _2 + i\epsilon {\bf K} +
i\epsilon {\bf K}2h \left (1-\frac {{\bf K}}{\eta}\right) \right )
\, . \label {Zn}
\end{eqnarray}
Now, we need to compute $\tilde Z_N(\theta)$ on each root of the
$n$ even string
\begin{equation}
\theta _1( \epsilon ', h')=\theta _1 + i \epsilon ' {\bf K}+i
\epsilon ' {\bf K} 2 h' \left (1-\frac {{\bf K}}{\eta} \right ) \,
, \quad \epsilon ' = \pm 1 \, , \quad h'=0,1,\ldots , k' \, ,
\label{roots}
\end{equation}
and sum up over all the roots (i.e. over both $\epsilon '$ and
$h'$). From the definition of counting function $\tilde
Z_N(\theta)$ we know (see discussion in Section 6 of \cite
{FRXYZ}) that the sum of the scattering phases (i.e. the terms
involving $\tilde \chi _{II}$, but not the forcing term) will
equalise exactly the scattering phase $i \ln S
_{2k'+2,2k+2}(\theta_{12}), \, \theta_{12}=\theta _1 -\theta_2$,
between the elliptic breathers $B_{2k'+2}(\theta _1)$ and
$B_{2k+2}(\theta _2)$. In other words, we can interpret on a
physical ground $S_{2k'+2,2k+2}(\theta_{12})$ as the scattering
amplitude. Technicaly, we shall pay attention to the positions of
the roots (\ref{roots}) in the second analyticity strip of $\tilde
\chi $, i.e. $\epsilon ' \, {\mbox {Im}}\, \theta _1( \epsilon ',
h')>2{\bf K}^2/\eta -2{\bf K}$. Therefore, we consider the second
determination of the functions
$\tilde\chi_{II}^{(\epsilon)}(\theta)$ in (\ref {Zn}): this
entails that we must apply again (\ref{chi2}), now with new input
$\tilde \chi _{II}^{(\epsilon)}(\theta)$ in the r.h.s.. We obtain
the new functions $\tilde \chi _{II}^{(\epsilon ',
\epsilon)}(\theta)_{II}$
\begin{equation}
\tilde \chi _{II}^{(\epsilon ',\epsilon)}(\theta)_{II}=i\ln \left
[-xq^{\frac {\epsilon+\epsilon '}{2}}\frac {\Theta _{p^2}\left
(x^{-1}q^{1-\frac {\epsilon+\epsilon '}{2}}\right )\Theta
_{p^2}\left (xq^{1+\frac {\epsilon+\epsilon '}{2}}p\right )} {
\Theta _{p^2}\left (x^{-1}q^{1-\frac {\epsilon+\epsilon
'}{2}}p\right )\Theta _{p^2}\left (xq^{1+\frac {\epsilon+\epsilon
'}{2}}\right ) } \right ] \, , \label {chiee}
\end{equation}
or explicitly
\begin{equation}
\tilde \chi _{II}^{(+,+)}(\theta)_{II}=i\ln \left [-xq\frac
{\Theta _{p^2}(x^{-1})\Theta _{p^2}(xq^2p)} {  \Theta
_{p^2}(x^{-1}p)\Theta _{p^2}(xq^2) } \right ] \, ,
\end{equation}
\begin{equation}
\tilde \chi _{II}^{(-,-)}(\theta)_{II}=i\ln \left [-xq^{-1}\frac
{\Theta _{p^2}(xp)\Theta _{p^2}(x^{-1}q^2)} {  \Theta
_{p^2}(x)\Theta _{p^2}(x^{-1}q^2p) } \right ] \, ,
\end{equation}
\begin{equation}
\tilde \chi _{II}^{(-,+)}(\theta)_{II}=\tilde \chi
_{II}^{(+,-)}(\theta)_{II}=i\ln \left [-x \frac {\Theta
_{p^2}(xqp)\Theta _{p^2}(x^{-1}q)} {  \Theta
_{p^2}(x^{-1}qp)\Theta _{p^2}(xq) } \right ] \, .
\end{equation}
Hence, the scattering factor to compute reads
\begin{eqnarray}
&&S _{2k'+2,2k+2}(\theta _{12})= \\
&=& {\mbox {exp}}\left [ i \sum _{h'=0}^{k'} \sum _{h=0}^k \sum
_{\epsilon '=\pm} \sum _{\epsilon =\pm} \tilde \chi _{II}
^{(\epsilon ', \epsilon)}\left ( \theta _{12}+i(\epsilon
'+\epsilon){\bf K}+i(\epsilon ' h'+\epsilon h )2 {\bf K}\left
(1-\frac {{\bf K}}{\eta}\right ) \right )_{II} \right ] \, .
\nonumber
\end{eqnarray}
and we can use the explicit expression (\ref{chiee}) to write down
\begin{equation}
S _{2k'+2,2k+2}(\theta _{12})=\prod _{h'=0}^{k'} \prod _{h=0}^k
\prod _{\epsilon '=\pm} \prod _{\epsilon =\pm} \left [ -x \frac
{\Theta _{p^2}(x^{-1}p^{f-1}q^{1-f})\Theta
_{p^2}(xp^{-f-1}q^{f+1}p)} {\Theta _{p^2}(xp^{-f-1}q^{1+f})\Theta
_{p^2}(x^{-1}p^{f-1}q^{1-f}p)} \right ]  \, ,
\end{equation}
with the shorthand $f=\frac {\epsilon +\epsilon '}{2}+\epsilon h +
\epsilon ' h'$ and where, with a little abuse of notation, we
defined again $x$ as
\begin{equation}
x={\mbox {exp}}\left (-\frac {2i\pi \eta \theta _{12}}{{\bf K}{\bf
K}'}\right ) \, .
\end{equation}
This expression may be rearranged as
\begin{equation}
S _{2k'+2,2k+2}(\theta _{12})=\prod _{h'=0}^{k'} \prod _{h=0}^k
\prod _{\epsilon '=\pm} \prod _{\epsilon =\pm} \left [
p^{f-1}q^{1-f} \frac {\Theta _{p^2}(xp^{1-f}q^{f-1})\Theta
_{p^2}(xp^{-f-1}q^{f+1}p)} {\Theta _{p^2}(xp^{-f-1}q^{1+f})\Theta
_{p^2}(xp^{1-f}q^{f-1}p)} \right ]  \, ,
\end{equation}
and, after performing the product over $\epsilon $, $\epsilon '$,
as
\begin{equation}
S _{2k'+2,2k+2}(\theta _{12})=\prod _{h'=-k'}^{k'+1} \prod
_{h=-k-1}^k \left [ p^{-1}q \frac {\Theta
_{p^2}(xp^{1-h-h'}q^{h+h'-1})\Theta
_{p^2}(xp^{-h-h'-1}q^{h+h'+1}p)} {\Theta
_{p^2}(xp^{-h-h'-1}q^{1+h+h'})\Theta
_{p^2}(xp^{1-h-h'}q^{h+h'-1}p)} \right ]  \, .
\end{equation}
Now, the product on $h'$ gives
\begin{eqnarray}
&&S _{2k'+2,2k+2}(\theta _{12})=\left (\frac
{q}{p}\right)^{(2k'+2)(2k+2)}\prod _{h=-k-1}^{k}
  \frac {\Theta _{p^2}(xp^{1+k'-h}q^{-k'+h-1})}{\Theta
_{p^2}(xp^{-k'-h-2}q^{2+k'+h})} \cdot \nonumber  \\
  && \cdot \frac {\Theta _{p^2}(xp^{k'-h}q^{-k'+h})
 \Theta _{p^2}(xp^{-k'-h-2}q^{k'+h+2}p)\Theta
_{p^2}(xp^{-k'-h-1}q^{k'+h+1}p)} {\Theta
_{p^2}(xp^{-k'-h-1}q^{1+k'+h})\Theta
_{p^2}(xp^{1+k'-h}q^{-k'+h-1}p)\Theta
_{p^2}(xp^{k'-h}q^{-k'+h}p)}=
  \nonumber \\
&&=\left (\frac {q}{p}\right)^{(2k'+2)(2k+2)} \frac {\Theta
_{p^2}(xp^{k'-k}q^{k-k'}) \Theta _{p^2}(xp^{k+k'+2}q^{-k-k'-2})
\Theta _{p^2}(xp^{-k-k'-2}q^{k+k'+2}p) }{\Theta
_{p^2}(xp^{-k-k'-2}q^{k+k'+2}) \Theta
_{p^2}(xp^{k-k'}q^{k'-k})\Theta _{p^2}(xp^{k+k'+2}q^{-k-k'-2}p)}
\cdot \nonumber \\
&& \cdot \frac {\Theta _{p^2}(xp^{k-k'}q^{k'-k}p)}{\Theta
_{p^2}(xp^{k'-k}q^{k-k'}p)}
 \prod _{h=-k-1}^{k-1} \frac {\Theta _{p^2}^2
(xp^{k'-h}q^{h-k'}) \Theta _{p^2}^2
(xp^{-h-k'-2}q^{h+k'+2}p)}{\Theta _{p^2}^2
(xp^{-h-k'-2}q^{h+k'+2}) \Theta _{p^2}^2 (xp^{-h+k'}q^{h-k'}p) }
\, .
\end{eqnarray}
Eventually, we can reach the final form after performing algebraic
manipulations in order to highlight the exchange symmetry
$x\rightarrow x^{-1}$ and with the positions $2k'+2=n, 2k+2=m$
\begin{eqnarray}
S_{n,m}(\theta _{12})&=&x^{2}\frac {\Theta _{p^2} (x^{-1}q^{\frac
{n-m}{2}}p^{\frac {m-n}{2}})\Theta _{p^2} (xp^{1+\frac {m-n}{2}}
q^{\frac {n-m}{2}}) \Theta _{p^2} (x^{-1}q^{\frac
{m+n}{2}}p^{-\frac {m+n}{2}})} {\Theta _{p^2} (x q^{\frac
{n-m}{2}}p^{\frac {m-n}{2}})\Theta _{p^2} (x^{-1}p^{1+\frac
{m-n}{2}} q^{\frac {n-m}{2}}) \Theta _{p^2} (xq^{\frac
{m+n}{2}}p^{-\frac {m+n}{2}})} \cdot
\nonumber \\
&& \label{Snm0} \\
&\cdot & \frac {\Theta _{p^2} (xp^{1-\frac {n+m}{2}} q^{\frac
{m+n}{2}})}{\Theta _{p^2} (x^{-1}p^{1-\frac {n+m}{2}} q^{\frac
{m+n}{2}})} \prod _{l=1+\frac {n-m}{2}}^{\frac {m+n}{2}-1} \left [
x^2\frac {\Theta _{p^2} ^2 (x^{-1}q^lp^{-l})\Theta _{p^2}^2(x
p^{1-l}q^l)} {\Theta _{p^2} ^2 (x q^lp^{-l})\Theta _{p^2}^2(x^{-1}
p^{1-l}q^l)} \right ]  \, . \nonumber
\end{eqnarray}
Albeit this formula has been obtained for both $n$ and $m$ even,
it is easy but lengthy, by following the same path, to show that
indeed formula (\ref{Snm0}) gives the scattering factor between
the general $n$-th and the $m$-th breather. However, two
precautions should be observed when dealing with (\ref{Snm0}). The
first one is that the product over $l$ is present only if $m\geq
2$; the second one concerns that this product runs over the
integers if $n$ and $m$ have the same parity, whereas over the
half-integers if $n$ and $m$ have different parities. We end this
part by noticing two simple properties of (\ref{Snm0})
\begin{equation}
S_{n,m}(\theta )S_{n,m}(-\theta )=1 \, , \quad S_{n,m}(\theta
)=S_{m,n}(\theta ) \, .
\end{equation}

\medskip

{\bf Remark 1}: The scattering factors between the elliptic
breathers, (\ref {Snm0}), are not independent. More precisely,
they satisfy the decomposition
\begin{eqnarray}
&&S_{n+m,n'+m'}\left [\theta -\left (\frac {i{\bf K}}{2}-\frac
{i{\bf
    K}^2}{2\eta}\right )(n-m-n'+m') \right ]= \nonumber  \\
&=& S_{n,m'} \left [\theta  -\left (\frac {i{\bf K}}{2}-\frac
{i{\bf
    K}^2}{2\eta}\right )(n+m+n'+m')  \right ] \cdot \nonumber  \\
&\cdot & S_{m,m'} \left [\theta  -\left (\frac {i{\bf K}}{2}-\frac
{i{\bf
    K}^2}{2\eta}\right )(-n-m+n'+m') \right ] \cdot \label {fus0}  \\
&\cdot & S_{n,n'} \left [\theta  -\left (\frac {i{\bf K}}{2}-\frac
{i{\bf
    K}^2}{2\eta}\right )(n+m-n'-m') \right ] \cdot \nonumber  \\
&\cdot & S_{m,n'} \left [\theta  -\left (\frac {i{\bf K}}{2}-\frac
{i{\bf
    K}^2}{2\eta}\right )(-n-m-n'-m') \right ] \nonumber \, .
\end{eqnarray}
In particular, this links the general scattering amplitude
$S_{n,n}$ to that of the fundamental scalar particle $S_{1,1}$
(or, in mathematical language, to the DVA structure function):
\begin{equation}
S_{n,n}(\theta)=\prod _{l=1-n}^{n-1}{S_{1,1}}^{n-|l|} \left [
\theta - \left ( 2i{\bf K}-\frac {2i{\bf K}^2}{\eta} \right)l
\right ]  \, . \label {fus1}
\end{equation}

{\bf Remark 2}: Let $B_n(\theta)$ be the Zamolodchikov-Faddeev
generator of the $n$-th breather with rapidity $\theta$ and
satisfy the exchange algebra
\begin{equation}
B_n(\theta _1)B_m(\theta _2)=S_{n,m}(\theta _1-\theta _2)
B_m(\theta _2)B_n(\theta _1) \, . \label{FZ}
\end{equation}
The relation (\ref{fus0}) is consistent with the following
connection between the creation operator $B_{n+m}(\theta)$ and the
operators $B_n(\theta)$ and $B_m(\theta)$:
\begin{eqnarray}
&&B_{n+m}\left [\theta+\left (\frac {i{\bf K}}{2}-\frac {i{\bf
    K}^2}{2\eta}\right )(m-n)\right ]=B_{m}\left [\theta+\left
    (\frac {i{\bf K}}{2}-\frac {i{\bf K}^2}{2\eta}\right )(n+m)\right] \cdot \nonumber \\
&\cdot &  B_{n}\left [\theta-\left (\frac {i{\bf K}}{2}-\frac
{i{\bf K}^2}{2\eta} \right) (n+m)\right ] \, . \label {fus3}
\end{eqnarray}
Consecutive applications of this property allows us to write the
generic breather generator $B_n$ as a product of fundamental
scalar generators $B_1$. This fact becomes important as soon as we
realise \cite{FRXYZ} that $B_1(\theta)$ coincide with the mode
generator of DVA, $T(z)$, which enjoys a particular structure
\cite{SKAO}. Hopefully, this should clarify the physical r\^ole
played by DVA. Finalising the parallelism, (\ref{fus3}) is the
elliptic deformation of the analogous relation in the sine-Gordon
theory (see footnote 9 in the seminal paper \cite {ZZ}).

\subsection{Recovering the trigonometric limit}
Let the trigonometric limit be defined by ${\bf K}'\rightarrow
\infty$ or, equivalently, by ${\bf K}\rightarrow \pi /2 $. In this
respect a key limit is
\begin{equation}
\frac {\Theta _{p^2}(x^{-1}q^lp^{-l})}{\Theta
_{p^2}(x^{-1}q^lp^{1-l})} \rightarrow \frac {1}{i}\tanh \left [
\frac {\theta}{2}+\frac {i\pi b}{2}l \right ] \, , \label{trilim}
\end{equation}
where we have introduced the quantity $b$ defined by $\eta$, and
then, thanks to the nature of that limit, connected to the
sine-Gordon coupling constant (see section 2) as follows:
\begin{equation}
\eta = \frac {\pi}{2(b+1)} \, , \quad b=\frac {\beta ^2}{1-\beta
^2} \, .
\end{equation}
Application of the limit (\ref{trilim}) to the scattering factor
(\ref{Snm0}) gives for us
\begin{eqnarray}
S_{n,m}(\theta)&\rightarrow &\frac {\tanh \left [ \frac
{\theta}{2}+\frac {i\pi b}{4}(n-m) \right ] \tanh \left [ \frac
{\theta}{2}+\frac {i\pi b}{4}(m+n) \right ] }{\tanh \left [ \frac
{\theta}{2}-\frac {i\pi b}{4}(n-m) \right ] \tanh \left [ \frac
{\theta}{2}-\frac {i\pi b}{4}(m+n) \right ] }\prod _{l=1+\frac
{n-m}{2}}^{\frac {m+n}{2}-1} \frac {\tanh ^2 \left ( \frac
{\theta}{2}+\frac {i\pi b}{2}l \right )} {\tanh ^2 \left ( \frac
{\theta}{2}-\frac {i\pi b}{2}l \right )}
\nonumber \\
&=&\frac {\sinh \theta + i \sin \frac {\pi b}{2}(n-m)} {\sinh
\theta - i \sin \frac {\pi b}{2}(n-m)} \, \, \frac {\sinh \theta +
i \sin \frac {\pi b}{2}(m+n)} {\sinh \theta - i
\sin \frac {\pi b}{2}(m+n)} \cdot \nonumber \\
&\cdot & \prod _{l=1}^{m-1} \frac {\sin ^2 \left [ i\frac
{\theta}{2}-\frac {\pi b}{4}(n-m+2l) \right ]\cos ^2 \left [
i\frac {\theta}{2}+\frac {\pi b}{4}(n-m+2l) \right ]} {\sin ^2
\left [- i\frac {\theta}{2}-\frac {\pi b}{4}(n-m+2l) \right ]\cos
^2 \left [ -i\frac {\theta}{2}+\frac {\pi b}{4}(n-m+2l) \right ]}
\, . \label{trigS}
\end{eqnarray}
As expected, this limiting formula coincides with the scattering
amplitude between the $n$-th and the $m$-th breather in the
sine-Gordon model (i.e. formula (4.19) of \cite{ZZ}).

\section{Higher Order Deformed Virasoro Algebras}
\setcounter{equation}{0} Continuing the identification between the
fundamental scalar generator $B_1(\theta)$ and the DVA generator
$T(z)$, we can now define higher order Deformed Virasoro Algebras.
The order $n$ deformed Virasoro algebra is the central extension
of the Zamolodchikov-Faddeev algebra \footnote{Some authors prefer
to define the Zamolodchikov-Faddeev algebra with the central
extension: in this case it would simply coincide with a higher
order deformed Virasoro algebra.} of two breathers creators $B_n$,
i.e.
\begin{equation}
B_n(\theta _1)B_n(\theta _2)=S_{n,n}(\theta _1-\theta _2)
B_n(\theta _2)B_n(\theta _1) \, , \label{FZ1}
\end{equation}
where
\begin{equation}
S_{n,n}(\theta)=-x \frac {\Theta _{p^2} (x^{-1}q^{n}p^{-n})\Theta
_{p^2} (xp^{1-n} q^{n})}{\Theta _{p^2} (xq^{n}p^{-n})\Theta _{p^2}
(x^{-1}p^{1-n} q^{n})}  \prod _{l=1}^{n-1} \left [ x^{2} \frac
{\Theta _{p^2} ^2 (x^{-1}q^lp^{-l})\Theta _{p^2}^2(xp^{1-l}q^l)}
{\Theta _{p^2} ^2 (xq^lp^{-l})\Theta _{p^2}^2(x^{-1} p^{1-l}q^l)}
\right ] \, .
\end{equation}
When $n=1$, the algebra (\ref {FZ1}) coincides with the Deformed
Virasoro Algebra ${Vir}_{p,q}$ \cite {SKAO} without central
extension. In order to clarify the relation between the general
(\ref{FZ1}) (or, which is the same, the corresponding higher order
deformed Virasoro algebra) and the particular ${Vir}_{p,q}$, we
consider the semiclassical limit of (\ref{FZ1}) around the
particular values $q=p^h$ (or ${\bf K}=\eta h$), $h=1,2$,
characterised by the special property that at them
$S_{n,n}(\theta)=1$. From the algebraic perspective, this means
that we can define a Poisson bracket around each point for which
$q=p^h$. In this respect, the order $n$ Zamolodchikov-Faddeev
algebra (generated by $B_n$) turns out to be a quantisation of
such Poisson structure. Explicitly, if we measure the distance
through $\beta$ such that $q^{1-\frac {\beta}{2}}=p^h$, these
Poisson brackets are defined by
\begin{equation}
\{ B_n(\theta _1) \, , \, B_n(\theta _2) \} = \frac {\partial
}{\partial \beta } S_{n,n}(\theta _1 -\theta _2) |_{\beta =0}
B_n(\theta _1) B_n (\theta _2) \, ,
\end{equation}
where
\begin{eqnarray}
&& \frac {\partial }{\partial \beta } S_{n,n}(\theta_{12}) |_{\beta =0}= C(n,h)\ln p \left [ \frac {x^{-1}}{1-x^{-1}}-\frac {x}{1-x}+ \right . \\
&& \left . + \sum _{m=0}^{\infty} \left ( - \frac
{2x^{-1}p^{2m}}{1-x^{-1}p^{2m}}+\frac
{2x^{-1}p^{2m+1}}{1-x^{-1}p^{2m+1}}+\frac
{2xp^{2m}}{1-xp^{2m}}-\frac {2xp^{2m+1}}{1-xp^{2m+1}} \right )
\right ] \, , \nonumber
\end{eqnarray}
and besides
\begin{equation}
C(n,h)=\sum _{l=1-n}^{n-1}(-1)^{(1-h)l}(n-|l|)\frac
{h}{2}(-1)^{h+1} \Rightarrow C(n,1)=\frac {n^2}{2} \, , \, \,
C(n,2)=\frac {(-1)^n-1}{2} \, .
\end{equation}
Interpreting this relation in terms of modes (see for instance the
discussion in Section 5 of \cite {AFRS}), we remark that the
Poisson structure around $q=p^h$ is proportional to the
Frenkel-Reshetikhin Poisson bracket \cite {FrRe}, which defines
the classical Deformed Virasoro Algebra, without central
extension. Even this conclusion has led us to consider the
breather algebras as higher order versions of the fundamental
DVA\footnote{Notice that it is `fundamental' even in a physical
sense as it describes the scattering of the fundamental scalar
particle.}.

\section {Conclusive outlook}
In this paper we have continued the study of scattering factors
between bound states of the spin 1/2-XYZ model, commenced in \cite
{FRXYZ} and restricted there to the simplest one. Following the
method exposed in \cite{FRXYZ}, any scattering factor has been
extracted from the infrared limit of the NLIE describing the
pairwise collision. In the end, we have found the complete set of
scattering factors between bound states (elliptic breathers) of
the spin 1/2-XYZ chain (in the attractive regime). This physical
information has allowed us to re-interpret the braiding relations
between the corresponding scattering operators as a consistent
mathematical definition of a family of Deformed Virasoro Algebras,
parameterized by a positive integer $n$. Of course, at $n=1$ the
usual Deformed Virasoro Algebra \cite{SKAO} is recovered and $n$
ranges from this value to a maximum depending on the chain
parameters. And all the members of this family have been proved to
reproduce, in the classical limit, the Poisson bracket structure
discovered by Frenkel and Reshetikhin \cite{FrRe} (and defining
the classical Deformed Virasoro Algebra). For these reasons, these
algebras are all eligible to the name of higher order Deformed
Virasoro Algebras and generalise the seminal quantisation by
Shiraishi-Kubo-Awata-Odake \cite{SKAO}. Besides, the search for a
family of DVAs seems to be a quite intriguing problem, though we
cannot guaranty that we have found the amplest family neither the
only one. Nevertheless, these results should shed light on the
peculiar r\^ole of DVA in field and lattice theory.

In an incoming paper \cite{35} these results and others will be
used to highlight the non-relativistic nature of the thermodynamic
theory and the link between bound state energies and the poles of
the Zamolodchikov $Z_4$ S-matrix \cite{ZAM}.

\vspace {1cm}

{\bf Acknowledgements.}\  Discussions with L. Frappat, M. Jimbo,
J. Shiraishi, P. Sorba are kindly acknowledged. DF thanks
Leverhulme Trust (grant F/00224/G) and PRIN 2004 "Classical,
quantum, stochastic systems with an infinite number of degrees of
freedom" for financial support and the Theory Group in Bologna for
a very warm welcome. MR thanks JSPS for the Invitation Fellowship
for Research in Japan (Long-term) L04716, and Lapth (Annecy,
France) and Luc Frappat for kind hospitality and support during
the last stages of this work.

\end{document}